\documentclass[3p,times]{elsarticle}

\usepackage{amssymb}
 \usepackage{amsthm}
\usepackage{amsfonts,graphicx} 
\usepackage{caption}           
\usepackage{subcaption}         
\usepackage{float}             
\usepackage{amsmath}             

\usepackage[figuresright]{rotating}

\begin{document}

\begin{frontmatter}




\title{Solving the $\hat{q}$ puzzle with $x$ and scale dependence}


\author[label1]{Evan Bianchi}
\author[label2]{ Jacob Elledge}
\author[label1]{ Amit Kumar}
\author[label1]{ Abhijit Majumder}
\author[label3,label1]{ Guang-You Qin}
\author[label4]{ Chun Shen}

\address[label1]{Department of Physics and Astronomy, Wayne State University, Detroit, MI 48201}
\address[label2]{Department of Physics and Astronomy, Arizona State University, Tempe,  AZ 85287}
\address[label3]{Institute of Particle Physics and Key Laboratory of Quark and Lepton Physics (MOE), Central China Normal University, Wuhan, 430079, China}
\address[label4]{Department of Physics, Brookhaven National Laboratory, Upton NY 11973}

\begin{abstract}
We present here a possible resolution of the jet quenching parameter $\hat{q}$ puzzle by exploring the momentum fraction $x$ and scale  $Q^2$ dependence of $\hat{q}$, in addition to  temperature dependence. We explore the momentum broadening of the jet due to  Glauber gluon exchange with the quark-gluon plasma (QGP). This has led us to define the momentum fraction $x$ for the partons in the QGP. We also show here, for the first time, possible forms of the parton distribution function (PDF) of the QGP. The QGP-PDF as input in $\hat{q}$ has been used  to fit the PHENIX  and CMS data simultaneously. We observe that the scale evolution of the QGP-PDF and the energy of the jet are the missing ingredients responsible for the enhancement of $\hat{q}$ at the same temperature in RHIC collisions compared to LHC collisions. 
\end{abstract}


\end{frontmatter}


\section{Introduction}
\label{}
In ultra-relativistic nucleus-nucleus collisions, calculation of high ${p_{T}}$ leading hadron suppression provides an important tool to investigate the properties of the QGP medium. Such calculations require the factorization of  initial state effects, and final state effects  from hard parton-parton  scattering. This factorization is a well established for proton-proton collision due to work by Collins, Soper, Sterman $\sim$ \cite{CollinsSoperSterman}, and generally assumed to hold for the case of nucleus-nucleus collision as well. Initial state effects are characterized by universal parton distribution functions. The differential cross section for high ${p_{T}}$ parton production is calculable with the the framework of perturbative QCD. Final state effects are encoded in the calculation through  universal vacuum fragmentation functions, and soft jet transport coefficients such as $\hat{q}$ (The average transverse momentum broadening per unit length) in the dense QGP medium.

In this article we focus on  the so called $\hat{q}$ puzzle. We will show that there is an integrated $x$ and scale dependence in $\hat{q}$ which emerge through QGP patron distribution functions.  We will demonstrate that this puzzle can be explained without invoking any non-trivial temperature dependence in ${\hat{q}}/{T^3}$. 
\section{Jet quenching parameter $\hat{q}$ puzzle}
The jet quenching parameter $\hat{q}$ characterizes the average transverse momentum broadening ($k_{\perp}$) of the parton traveling through medium per unit length ($L$) given as
$
\hat{q} = \frac{k^{2}_{\perp}}{L}.
$
Existing  in-medium energy loss models parameterize $\hat{q}$ with the local temperature $T^3$, or energy density $\epsilon^{3/4}$, or entropy density of the medium \cite{JETCol}.  Such models have a free parameter that is estimated through fitting the experimental data to the full model calculation. These phenomenological studies indicate that the free parameter in $\hat{q}/{T^3}$ is larger at RHIC than at the LHC. This odd property is known as the jet quenching parameter $\hat{q}$ puzzle. So far, there have been a handful of attempts to explain the cause of this behavior, based on parameter dependencies within $\hat{q}/T^3$  \cite{JETCol,Gyulassy,Carlota}. Recent studies  by the authors of Ref. \cite{Carlota} reveal that the free parameter $\hat{q}/{T^3}$ is sensitive to center-of-mass energy of nucleus-nucleus collision rather than local temperature of the QGP medium. In contrast to this, the authors of Ref. \cite{Gyulassy} suggest that the free parameter $\hat{q}/{T^3}$ might have a non-trivial (upward cusp-like) temperature dependence. In this article, we propose that enhancement in $\hat{q}/{T^3}$ at RHIC can be explained by invoking an integrated $x$ dependence and a scale dependence.

\section{Resolution of the $\hat{q}$ puzzle}
We consider an analogous scenario where a jet produced from a hard virtual photon   with light-cone momentum $q = [\frac{-Q^2}{2q^{-}},q^{-},0,0]$ striking a nucleon $A$ in a nucleus and generating a hard quark, which then 
propagates with momentum $[\frac{-Q^2}{2q^{-}} +p^{+},q^{-},0,0]$ through the remainder of the nucleus. This jet interacts with a nucleon $B$ in the nucleus by exchanging a Glauber gluon ($k_{\perp} >> k^{+}, k^{-}$), and  emerges in the final state with a non-vanishing transverse momentum $\vec{l}_{q\perp}$ as shown in figure \ref{QHatDiagram}(a).

\subsection{Expression for $\hat{q}$ }
\begin{figure}[h!]
\centering 
\begin{subfigure}{0.35\textwidth}
  \includegraphics[width=\textwidth]{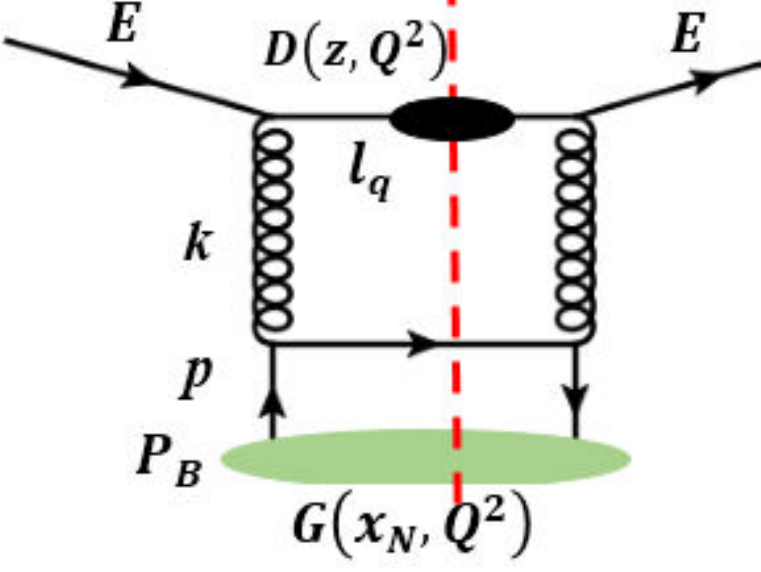}
\caption{}
\end{subfigure}
  \begin{subfigure}{0.27\textwidth}
   \includegraphics[width=\textwidth]{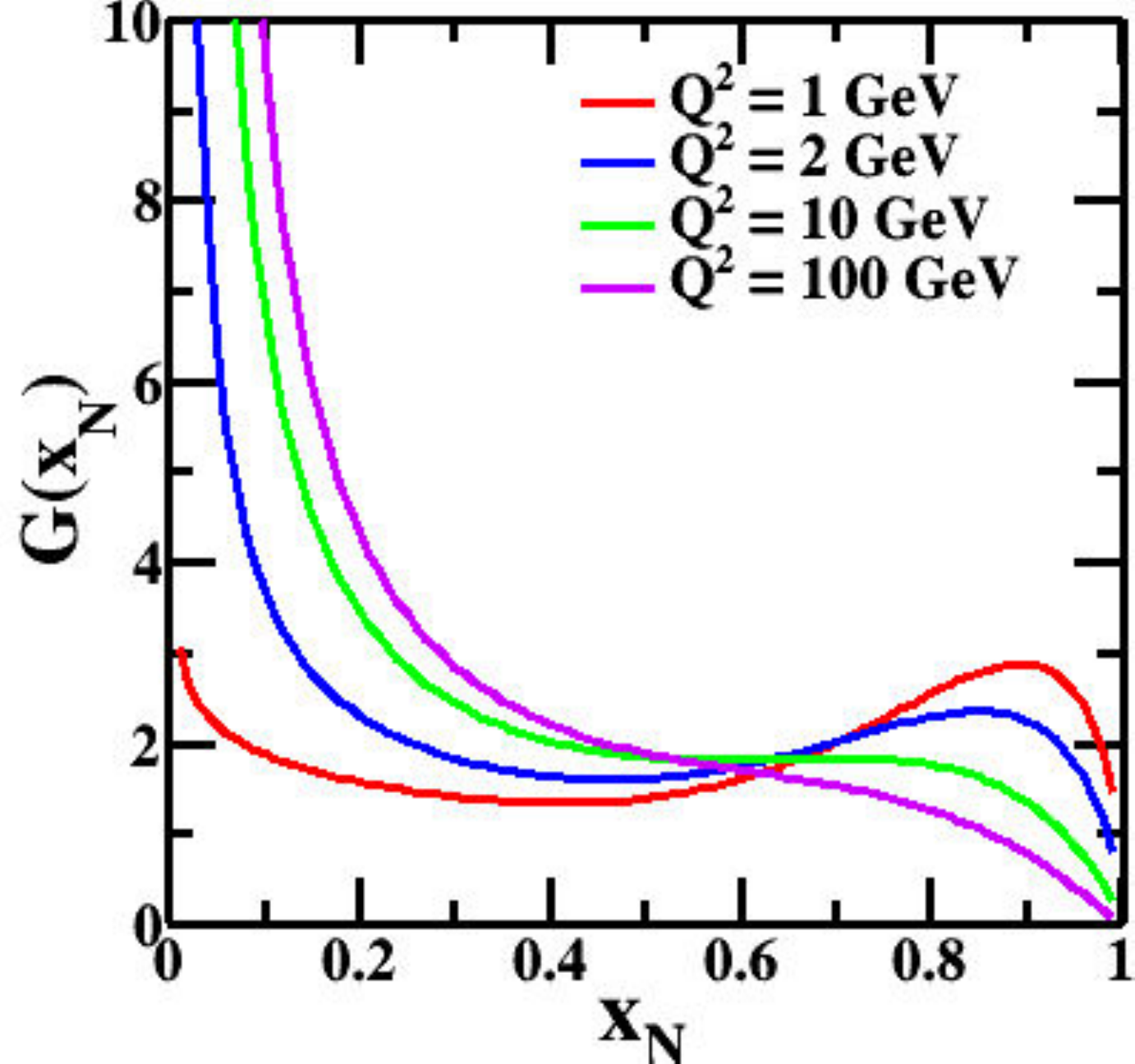}
\caption{}
    \end{subfigure}    
       \quad
      \begin{subfigure}{0.28\textwidth}
   \includegraphics[width=\textwidth]{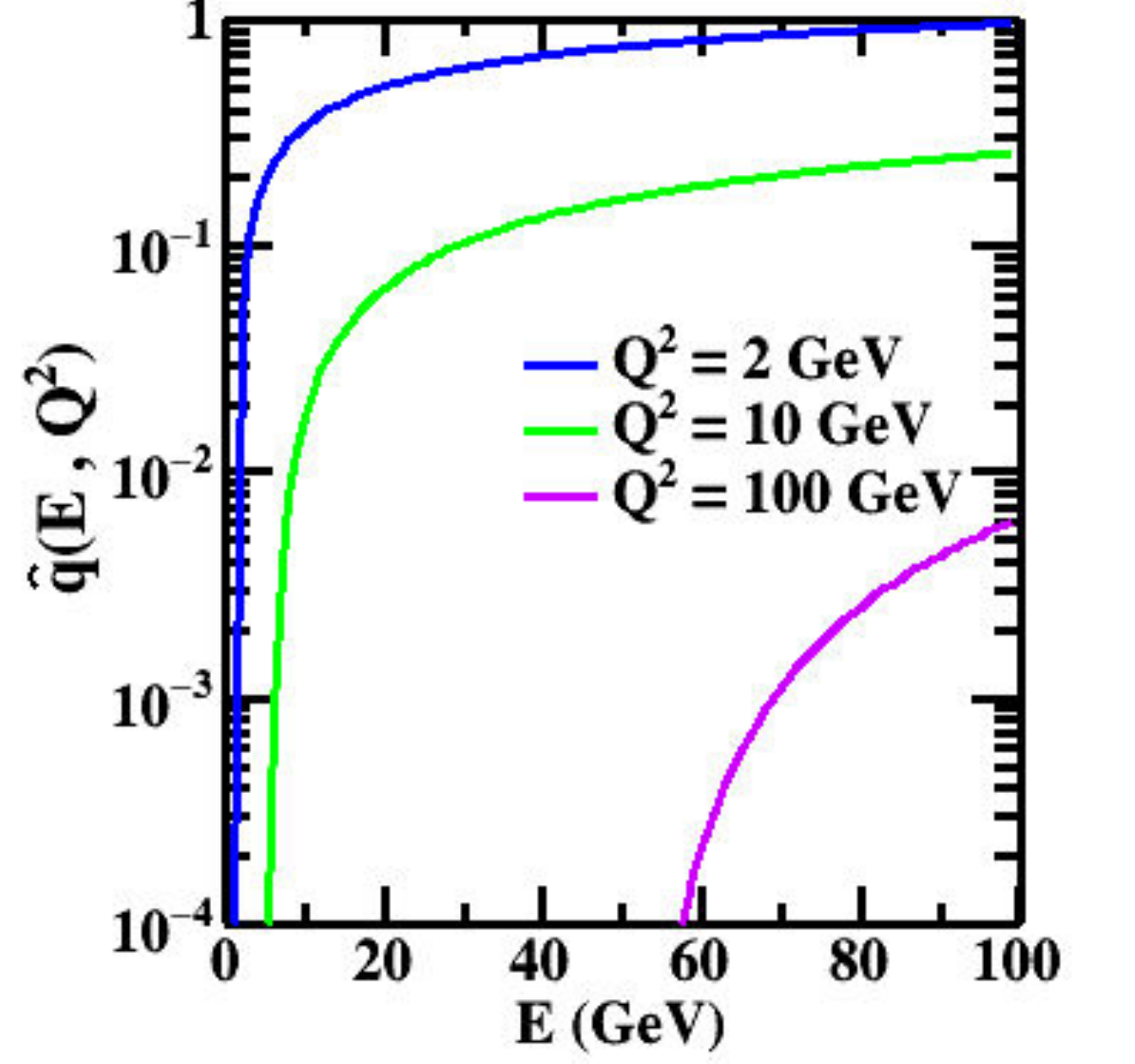}
   \caption{}
    \end{subfigure} 
    \caption{Evaluating $\hat{q}$. (a) Jet propagating through a nucleus interact via exchange of Glauber gluon emitted from a parton inside nucleon $B$. (b) DGLAP evolution of PDF. (c) Scale evolution of $\hat{q}$ in rest frame of nucleon $B$ for the process shown in figure \ref{QHatDiagram}(a) and PDF (figure \ref{QHatDiagram}(b)) as input.  }
    \label{QHatDiagram}
\end{figure}

We  calculate the hadronic tensor ($W^{\mu\nu}$) for the case of the virtual photon striking a single quark in one nucleon of a large nucleus, followed by the propagation of this quark through the nuclear medium undergoing a single rescattering on a nucleon down the path of the jet. The differential hadronic tensor is given as,
\begin{equation}
\begin{split}
\frac{dW^{\mu\nu}}{d^{2}l_{q\perp}} & = C^{A}_{P_AP_B}W^{\mu\nu}_{0}\left[\nabla^{2}_{l_{q\perp}} \delta^{2}(\vec{l}_{q\perp})  \right] \int^{L^{-}}_{0}dY^{-} \\
& \hspace{1cm} \times \left[   \frac{4 \pi^{2}\alpha_{s} }{2N_{c}} \int \frac{dy^{-}}{2\pi} \int \frac{d^{2}k_{\perp} d^{2}y_{\perp} }{(2\pi)^2} \exp^{-i\frac{k^{2}_{\perp}}{2q^{-}}y^{-} + i\vec{k}_{\perp}.\vec{y}_{\perp}} \times \left\langle P| F^{a+\alpha}(Y^{-} + y^{-}, \vec{y}_{\perp}) F^{a+}_{\alpha}(Y^{-}) |P  \right\rangle \right] \\
& =  C^{A}_{P_AP_B}W^{\mu\nu}_{0}\left[\nabla^{2}_{l_{q\perp}} \delta^{2}(\vec{l}_{q\perp})  \right] \int^{L^{-}}_{0}dY^{-} \hat{q}(Y^{-}),
\end{split}
\label{QHATEqu}
\end{equation}
where $W^{\mu\nu}_{0}$ represents hadronic tensor at leading order (for case with no rescattering of jet with nucleon). We have included a factor $C^{A}_{P_{A}P_{B}}$ to incorporate the probability to find a nucleon $A$ with momentum $P_A=[P^{+}_{A},0,0,0]$ in a nucleus,  and its correlation with the nucleon $B$ having momentum $P_B=[P^{+}_{B},0,0,0]$, down the path of the jet.  The quantity within square brackets (second line of Eq. $\sim$ \eqref{QHATEqu}) represents the desired quantity $\hat{q}$. We can see  from figure \ref{QHatDiagram}(a) that the parton which comes off nucleon $B$ has a parton distribution function, which in turn emits a Glauber gluon responsible for momentum broadening of the jet. We know that the nucleon PDF has a momentum fraction $x$ dependence and may also undergo evolution with scale, we can conclude that $\hat{q}$ (Eq. \eqref{QHATEqu}) also must have $x$ and scale dependence. We show nucleon PDFs for several different scales ($Q^2$) computed using Dokshitzer Gribov Lipatov Altarelli Parisi (DGLAP) evolution \cite{DGLAP} equation for an input nucleon PDF at $Q^2=1$ GeV (figure \ref{QHatDiagram}(b)). We have also computed   $\hat{q}$ (figure \ref{QHatDiagram}(c)) for the process shown in figure \ref{QHatDiagram}(a)  with input PDF for nucleon $B$ from figure \ref{QHatDiagram}(b).  We conclude that $\hat{q}$ is enhanced for small $Q^2$, and hence provides an explanation of the puzzle of  $\hat{q}$ being enhanced at RHIC than at LHC for the same temperature.

\subsection{Quark gluon plasma parton distribution function}
\begin{figure}[h!]
 \centering
 \begin{subfigure}{0.36\textwidth}
  \includegraphics[width=\textwidth]{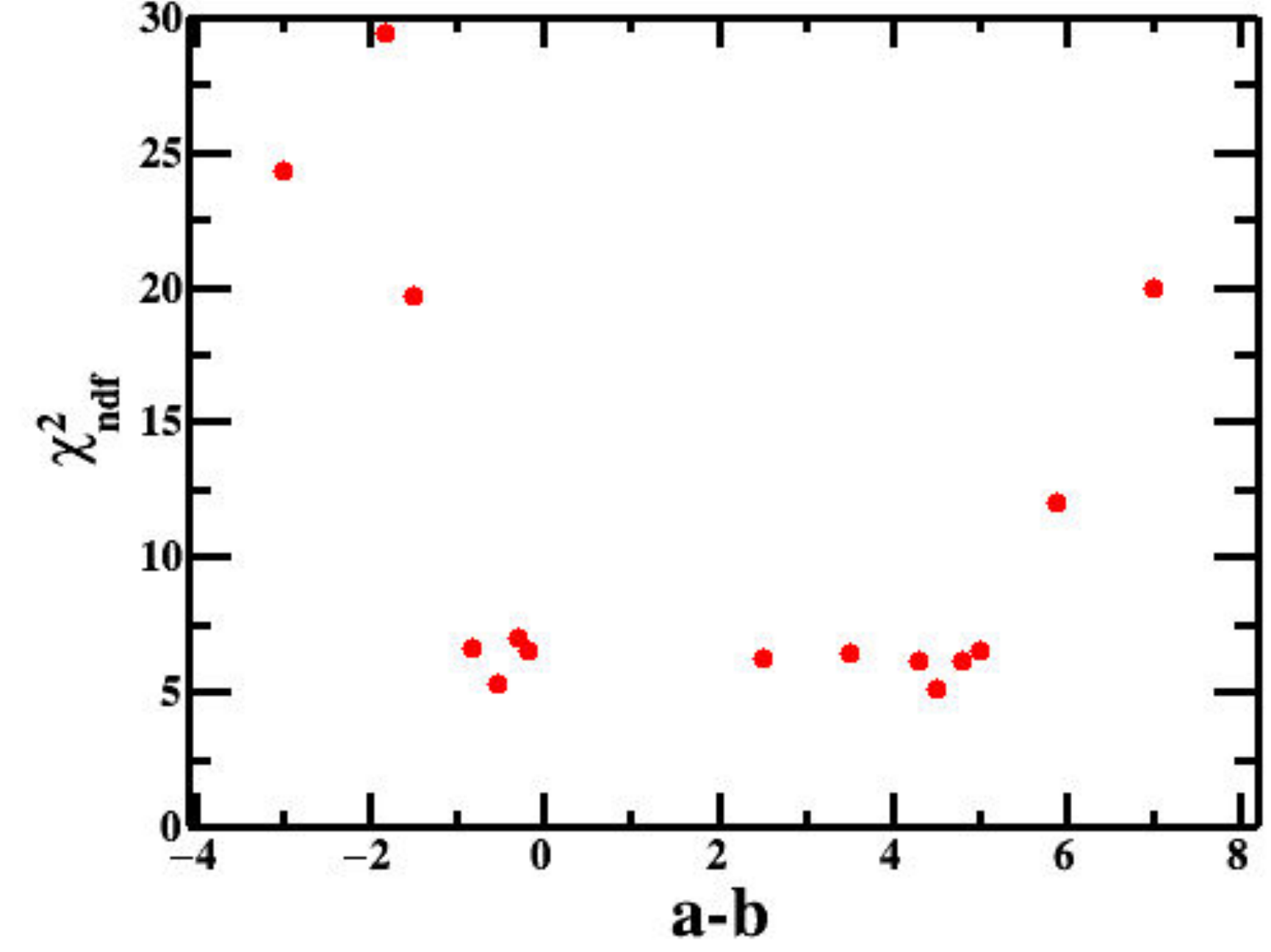}
  \caption{}
\end{subfigure}
      \begin{subfigure}{0.26\textwidth}
  \includegraphics[width=\textwidth]{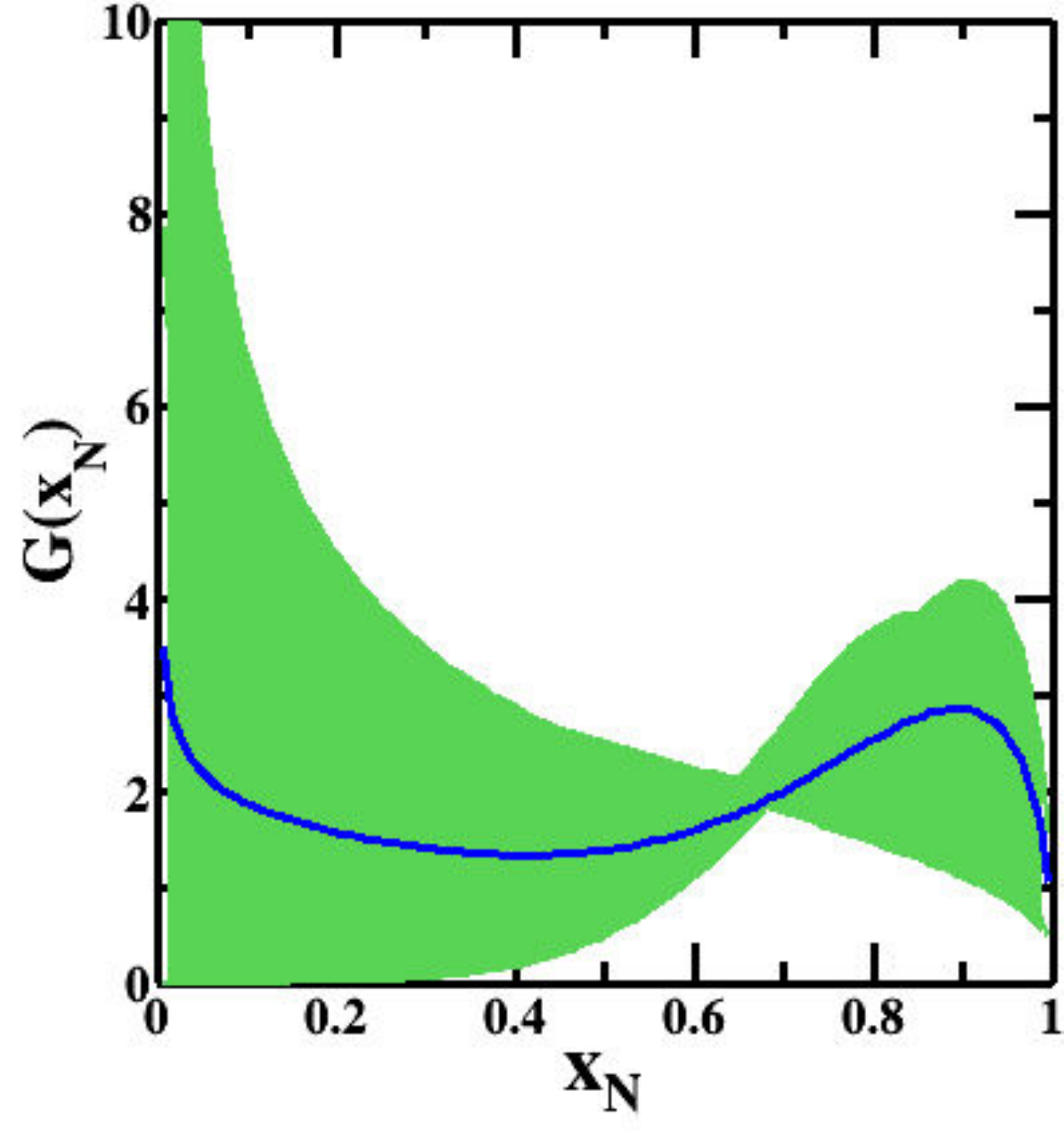}
\caption{}
      \end{subfigure}    
      \begin{subfigure}{0.26\textwidth}
  \includegraphics[width=\textwidth]{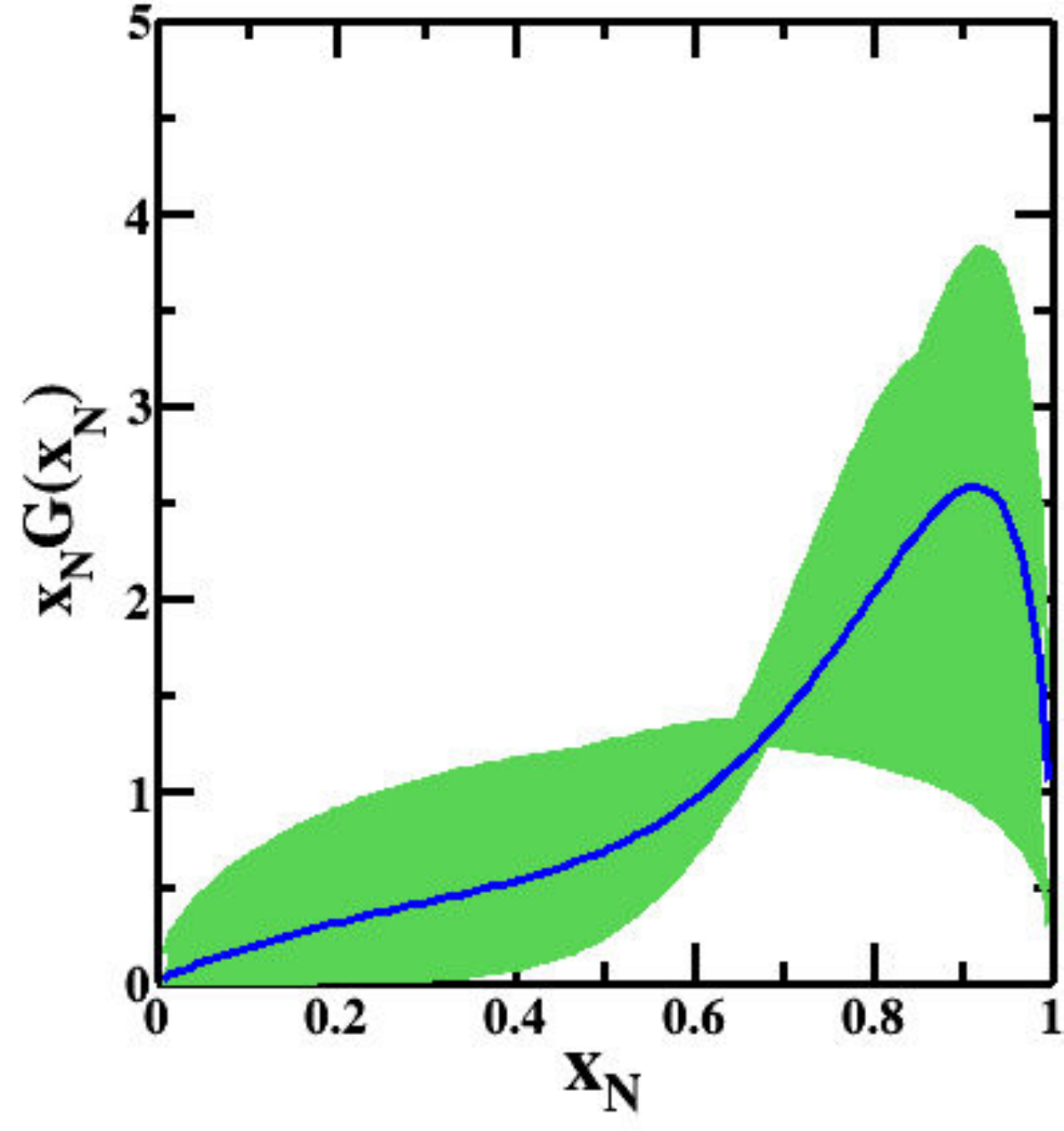}
\caption{}  
      \end{subfigure}    
      \caption{Quarks gluon plasma PDF: $G(x_{N},Q^2=1\mathrm{GeV}^2) = Nx_{N}^{a}(1-x_{N})^{b}$.(a) Combined chi-square of $R_{AA}$ (simultaneously at RHIC and LHC) for a given choices of fit parameters $a$ and $b$.(b) Possible form of the QGP PDF's. We have shown the calculation of $R_{AA}$ and $v_2$ in figure \ref{RAAV2} for the input QGP PDF shown in blue color. (c) The momentum fraction weighted PDF. .} 
      \label{ChiSquareQGP}
\end{figure}

We have discussed the dependence of $x$ and scale of $\hat{q}$ for the case of jet traveling through nucleus. We can generalize this argument to jets traveling through a QGP medium, if we could define momentum fraction $x$. We employ a concept of mass degree of  freedom ($m_{QGP}$). Since, $m_{QGP}$ is an unknown quantity, we use estimates from finite-temperature field theory which puts $m_{QGP} \sim gT $.   It implies that the momentum of a parton within a QGP degree of freedom is a temperature dependent quantity with an upper bound $gT \leq M_{N}$, where $M_{N}$ is mass of a nucleon (1 GeV).
We have calculated $\hat{q}$ in the rest frame of a QGP medium, and use $M_N$ to define the momentum fraction of a parton inside a QGP degree of freedom. We parametrize the input PDF in the calculation of $\hat{q}$ to the Feynman-Field  form \cite{FFPDF}:

\begin{equation}
G(x_{N},Q^2=1\mathrm{GeV}^2) = Nx_{N}^{a}(1-x_{N})^{b},
\end{equation}  
where $N, a$, and $b$ are fit parameters. We calculate the nuclear modification factor $R_{AA}$ using the multiple scattering, multiple emission higher-twist theory described in Ref. \cite{MajumderCShen}, together with $\hat{q}$ formalism introduced in this article. We vary parameters $N, a$, and $b$ to produce best fit to $R_{AA}$ data at PHENIX for centrality $0-10\%$, and CMS  for centrality $0-5\%$ simultaneously. With that choice of parameters we have shown the $R_{AA}$ for four different centralities   (top three panel in figure \ref{RAAV2}(a)). The QGP PDF used for best fit in this calculation is shown in blue color in figure \ref{ChiSquareQGP}(b, c). We have done the combined Chi-Square (Four different centralities for RHIC and LHC) analysis to estimate the allowed fit parameter space (figure \ref{ChiSquareQGP}(a)).  We show a band of QGP PDF's which reproduce the  $R_{AA}$ data at RHIC and LHC simultaneously with combined Chi-Square $\chi^2 _{dof} < 7.0$ (figure \ref{ChiSquareQGP}(b, c)). We observe that QGP PDF at scale $Q^2 =$ 1 GeV has a sea-like as well as narrow-valence like parton distribution.

\subsection{Calculation of observable ${R_{AA}}$ and ${v_2}$}
\begin{figure}[h!]
 \centering
      \begin{subfigure}{0.488\textwidth}
  \includegraphics[width=\textwidth]{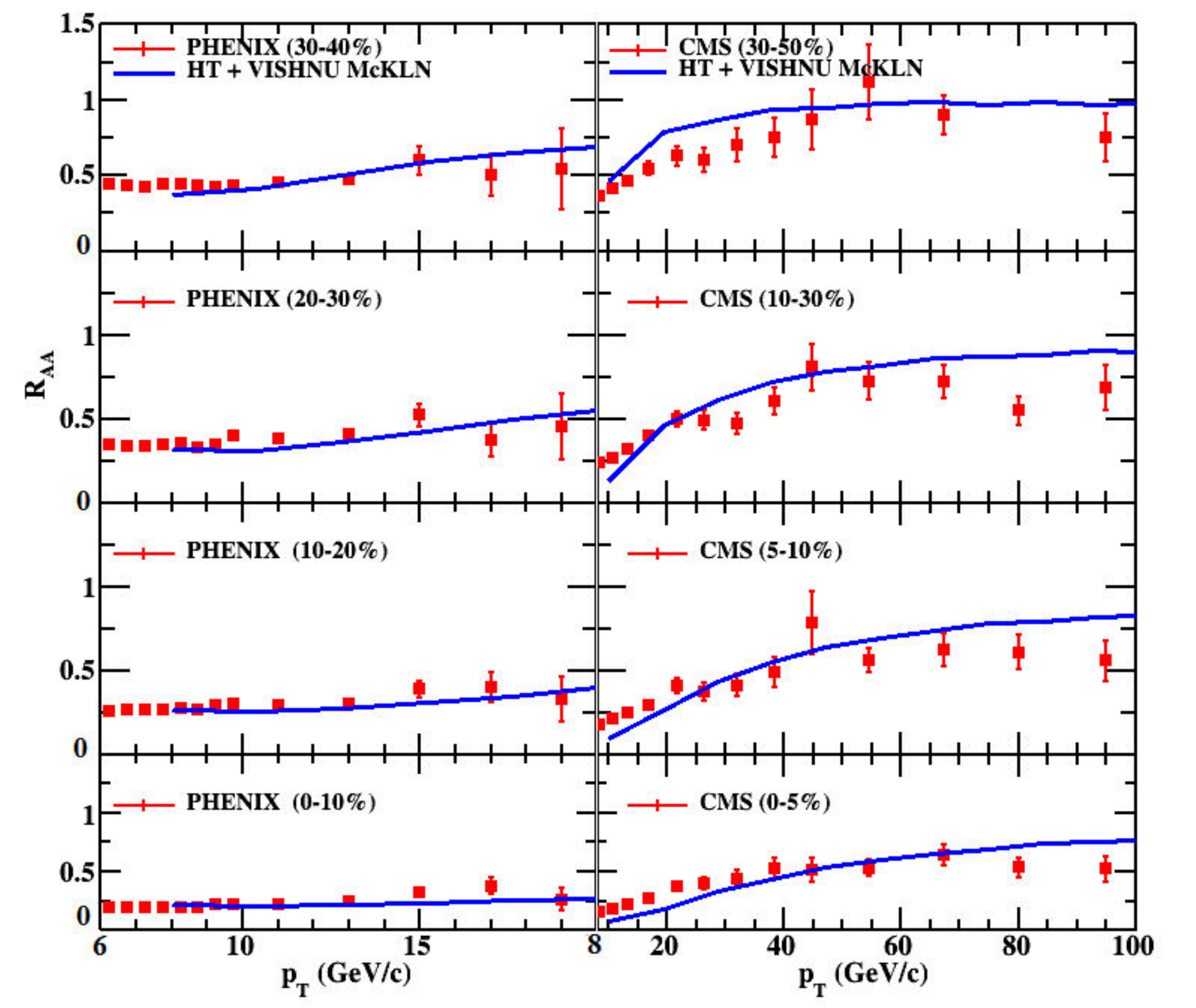}
  \caption{}
\end{subfigure}
      \begin{subfigure}{0.488\textwidth}
  \includegraphics[width=\textwidth]{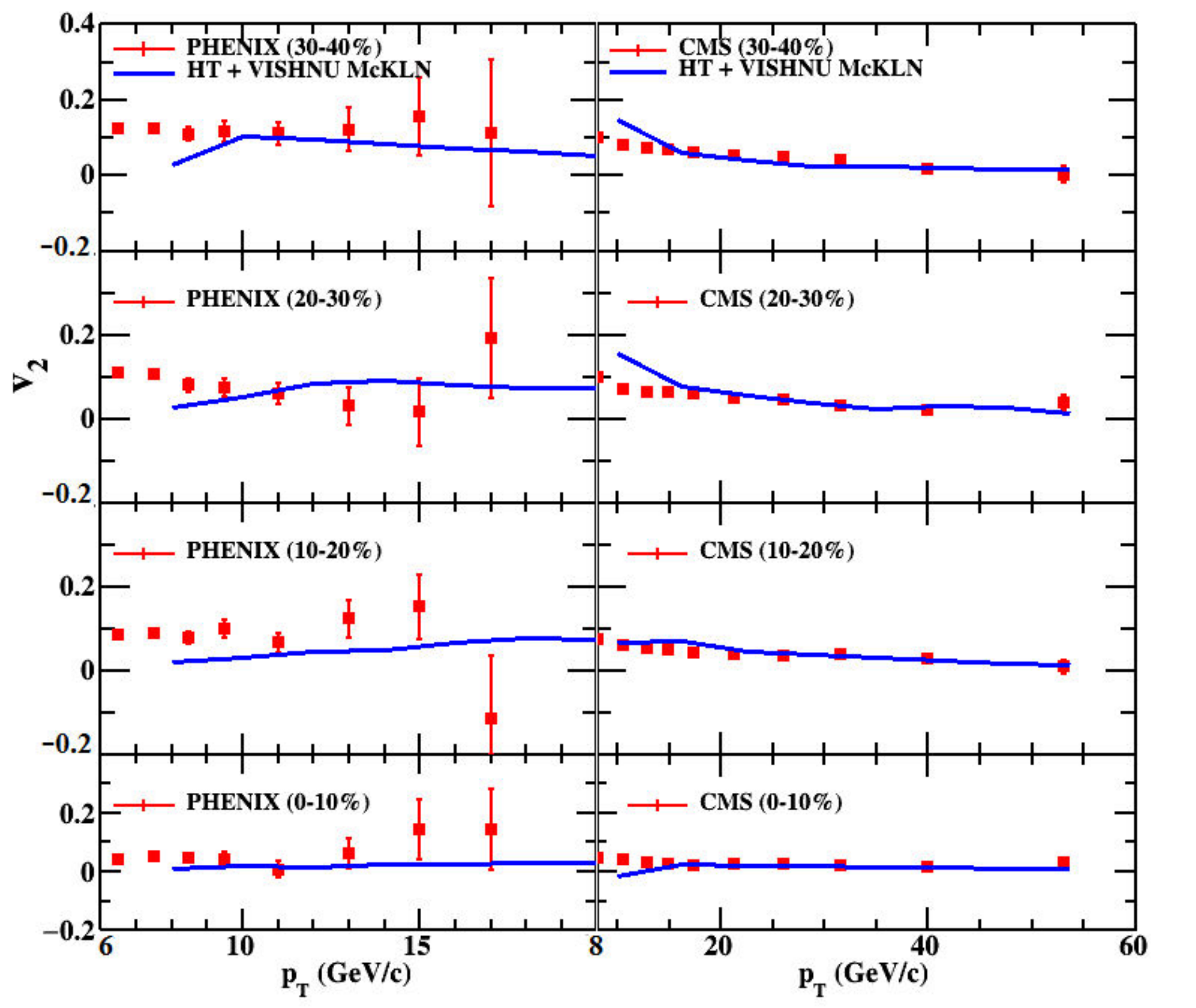}
    \caption{}
\end{subfigure}
    \caption{Higher-twist calculation having multiple emission and multiple scattering with the QGP PDF as input in $\hat{q}$. (a) Nuclear Modification factor $R_{AA}$. (b) Azimuthal anisotropy $v_{2}$.}
\label{RAAV2}
\end{figure}

Nuclear modification factor $R_{AA}$ is shown in figure \ref{RAAV2}(a). We have also calculated azimuthal anisotropy $v_2$ using QGP PDF as input (Blue curve in figure \ref{ChiSquareQGP}(b,c)). We show  $v_2$ results for two different collision energies for four different centralities (figure \ref{RAAV2}(b)). We see from plot in figure \ref{RAAV2}, a fair agreement between experiment and our full-model calculation. We point out that the calculation of the azimuthal anisotropy does not involve any further tuning of parameters. All parameters are set by the $R_{AA}$ calculation.
\section{Summary}
In this article we have successfully provided an explanation to the jet quenching parameter $\hat{q}$ puzzle. We have demonstrated that the $R_{AA}$ data at RHIC and LHC can be explained simultaneously, without two separate normalization of  $\hat{q}$, by inclusion of a scale evolution and an integrated $x$ dependence in the $\hat{q}$. The inherent scale dependence of the QGP PDF gives rise to the scale evolution of the $\hat{q}$. Based on chi-square fits to the $R_{AA}$ data, we also presented the possible forms of the QGP PDF. Furthermore, we have shown theoretical results for $R_{AA}$ and $v_2$ observables at RHIC and LHC energies which are in good agreement with the experimental data. 

This work was supported in parts by the NSF under grant numbers \rm{PHY-1207918}, and \rm{PHY-140853}, by the U.S. DOE , office of science, office of nuclear physics under grant number \rm{DE-SC0013460}, and 
by the NSFC under grant number 11375072.  




\bibliographystyle{elsarticle-num}
\bibliography{<your-bib-database>}



\end{document}